\documentclass[unsortedaddress,floatfix, twocolumn, aps,pra,review]{revtex4-1}
\usepackage{graphicx}
\usepackage{amsmath}
\usepackage{xcolor}
\begin{document}

%\preprint{}

\title{Detecting Macroscopic Indefiniteness of Cat States in Bosonic Interferometers}


\author{Shane P. Kelly}
\email[]{skell013@ucr.edu}
\affiliation{Theoretical Division, Los Alamos National Laboratory, Los Alamos, New Mexico 87545, USA}
\affiliation{Department of Physics and Astronomy, University of California Riverside, Riverside, California 92521, USA}

\author{Eddy Timmermans}
\affiliation{XCP-5, XCP Division, Los Alamos National Laboratory, Los Alamos, New Mexico 87545, USA}

\author{S.-W. Tsai}
\affiliation{Department of Physics and Astronomy, University of California Riverside, Riverside, California 92521, USA}


\date{\today}

\begin{abstract}

    The paradigm of Schr\"{o}dinger's cat illustrates how quantum states preclude the assignment of definite properties to a macroscopic object (realism).
    In this work we develop a method to investigate the indefiniteness of cat states using currently available cold atom technology.
    The method we propose uses the observation of a statistical distribution to demonstrate the macroscopic distinction between dead and alive states, and uses the determination of the interferometric sensitivity (Fisher information) to detect the indefiniteness of the cat's vital status.
    We show how combining the two observations can provide information about the structure of the quantum state without the need for full quantum state tomography, and propose a measure of the indefiniteness based on this structure.
    We test this method using a cat state proposed by Gordon and Savage [Phys. Rev. A 59, 4623 (1999)] which is dynamically produced from a coherent state.
    As a control, we consider a set of states produced using the same dynamical procedure acting on an initial thermal distribution. 
    Numerically simulating our proposed method, we show that as the temperature of this initial state is increased, the produced state undergoes a quantum to classical crossover where the indefiniteness of the cat's vital status is lost, while the macroscopic distinction between dead and alive states of the cat is maintained.
  
\end{abstract}

\maketitle

Superposition is at the heart of the many predictions made by quantum mechanics that clash with everyday intuition.
It allows for the possibility of an experiment in which we must conclude that some property of an object can not be prescribed a definite value before measurement.
Instead, this indefiniteness of a property must be modelled by a superposition of possible values and implies a statistical uncertainty that can not be reduced by obtaining more knowledge about the universe.
While plausible for microscopic properties, this possibility directly conflicts with our everyday intuition for macroscopic objects.
The characteristic example is the Schr\"{o}dinger's cat thought experiment\cite{schrodinger1935gegenwartige}{, where a cat ends up in a superposition of alive and dead by entangling with the decayed or excited state of a radioactive source}.

When investigating these macroscopic states in an experiment, we are naturally led to two questions: 1) How do we know the cat's life was an indefinite property before measurement? 2) How do we quantify the macroscopicity of the cat and thus, the extent to which it conflicts with our intuition about the macroscopic world?
The first question is answered by Leggett-Garg\cite{Leggett1985}, who constructed a set of inequalities on a set of different-time correlation functions that would only be {violated} if the cat was in an {indefinite} state at some intermediate time.
The second question has been answered by constructing measures of macroscopicity in two general ways\cite{RevModPhys.90.025004}: either by focusing on the structure of a macroscopic cat state\cite{Leggett1980,Leggett2002,Cirac2002,Mana2004,Korsbakken2007,Marquardt2008,Sekatski2014} or generalizing to any macroscopic quantum state\cite{Shimizu2002,Cavalcanti2006,Lee2011,Frowis2012a,Nimmrichter2013,Yadin2015,LAGHAOUT2015,Kwon2017}. For many of these measures, a state is declared macroscopic based on how the measures scale with the number, $N$, of constituent particles. The experimental observation of these measures often leads to a way to answer the first question\cite{Frowis2016,Cavalcanti2006,Kwon2017}.

In this paper we will work with a measure that is a combination of the one proposed by Leggett\cite{Leggett1980,Leggett2002} and the one proposed by Fr{\"o}wis and D{\"u}r\cite{Frowis2012a}.
The measure proposed by Leggett is quantified by two numbers: the extensive difference, $\Lambda$, which is the difference of the expectation value for some observable $A$ between the dead and alive states of the cat, and the disconnectivity, a quantity based on the entanglement entropy.
The extensive difference describes how macroscopically different the dead and alive cats are, while the disconnectivity quantifies how indefinite the vital status of the cat is.
The measure of Fr{\"o}wis and D{\"u}r\cite{Frowis2012a}, $N_{eff}$, is applicable to general quantum states and is based on the experimentally quantifiable, quantum Fisher information (QFI).
The QFI has been interpreted as a measure of entanglement\cite{Hyllus2012}, and has stimulated a variety of work studying this type of entanglement\cite{Smerzi2018,Hyllus2012,Frowis2012a, Lapert2012, Oberthaler2010,Gross2008, Strobel2014,Mirkhalaf2018}.
{The QFI has also been shown to be connected with the resource theory of coherence\cite{tan2018} and to be the maximum quantifier for the resource theory of quantum invasiveness\cite{moreira2019}.}
Inspired by the measure of Fr{\"o}wis and D{\"u}r, and by recent insights relating the QFI to the convex-roof of uncertainty\cite{Toth2013,Yu2013}(see Section II), we replace the disconnectivity in Leggett's measure by a function of the QFI and statistical variance.

This choice is further motivated by the fact that the extensive difference and the QFI are both experimentally accessible in bosonic interferometer experiments.
The kind of bosonic interferometer experiments discussed here\cite{Smerzi2018,Muessel2014,Strobel2014,Zibold2010,Riedel2010,Albiez2005a, Gross2006, Julia-Diaz2012, Baumann2011,Berg2015,Rosi2015,Debs2011,Muessel2014, Berg2015,Rosi2015,Debs2011} can be understood as a way to estimate a phase encoded onto a macroscopic spin by a projective measurement.
The maximum sensitivity of the interferometer to the encoded phase is given by the classical Fisher information (CFI) via the Cramer-Rao bound\cite{cramer1945,rao1945} and is restricted by the phase encoding method and the chosen projective measurement.
The QFI quantifies the sensitivity of the interferometer when the best projective measurement is used and is bounded from below by the CFI.
The CFI and other measures of sensitivity can be measured by experiments\cite{Strobel2014,Oberthaler2016} and many proposals exist to optimize the bound the CFI puts on the QFI\cite{Mirkhalaf2018,Garttner2018,Frowis2016}.
The extensive difference can also be obtained in an experiment from the counting statistics of a single-particle observable\cite{Oberthaler2016,Strobel2014}.

Various types of macroscopic states have been produced in these systems, ranging from squeezed states\cite{Oberthaler2010,Gross2008} to non-Gaussian entangled states\cite{Strobel2014}.
There also exists many proposals to create macroscopic cat states in bosonic interferometers\cite{Lapert2012,Hatomura2018,Mahmud2005,Micheli2003,Lau2014a,Carr2010a,Huang2006,Gordon1999,Garcia-March2011,Carr2010}.
In this article we work with a cat state first proposed by Gordon and Savage\cite{Gordon1999}.
The method for creating this state can be understood from the classical dynamics of the effective collective spin. As we explain in Section I, the classical dynamics exhibit two different kinds of trajectories separated in phase space by the separatrix. As pointed out by Micheli \textit{et al.}\cite{Micheli2003}, the cat state is prepared by creating an initial coherent state with a Wigner distribution that spans the phase space region crossing the separatrix.
The quantum dynamics then separates the components from either side of the separatrix into the macroscopically distinct alive (free oscillation) and dead (self-trapping) components of the cat.
They prove this by semi-classically evolving the Wigner function and finding it produces a double peak distribution in the z-component of the macroscopic spin. 

Similar arguments can be applied to mixed states, and we show that initial thermal distributions also evolve into a double peak state.
We show that the higher the temperature the less indefiniteness the state displays, and we describe how an experimenter can observe this transition.
These high temperature states are particularly appealing because, despite increasing the temperature, it is still possible to identify the dead and alive states of the cat.
Thus, as temperature increases, the vital status of the cat becomes definite before the distinction between dead and alive is loss.

Previous work has suggested the detection of indefinite properties for similar states by using generalized Leggett-Garg inequalities\cite{Rosales2018} or observation of many-body correlation functions\cite{Opanchuk2016}, but these methods rely on experimental tools that have yet to be implemented.
In this article, we study the possibility of currently available cold atom technology to experimentally detect the macroscopic indefiniteness of these cat states, and distinguish them from the classical uncertainty of the high temperature mixed states.
The method we propose uses the observation of a statistical distribution to demonstrate the macroscopic distinction (extensive difference) between dead and alive states and uses the interferometric sensitivity (QFI) to detect the indefinite  vital status of the cat.
We show how these two types of observations provide information about the nature of the possible pure states which make up the density matrix, and how this information is useful in observing the crossover from a cat {that is} in a superposition of dead and alive to a cat that is either dead or alive.
{Next, we numerically simulate the method {for the Gordon and Savage cat state} and demonstrate the quantum to classical crossover.
    { Inspired by the Schr\"{o}dinger's cat thought experiment, we conclude by considering a cat state which is entangled with an auxiliary qubit (representing the radioactive source) and show that such a quantum to classical crossover is controlled by the strength of entanglement with the auxiliary qubit.}

\section{Interferometers, cat states and double peak mixed states}

Interferometry in Bose Einstein condensates has led to new measurement techniques for magnetic fields\cite{Muessel2014}, gravitational fields\cite{Rosi2015,Debs2011} and rotational motion\cite{Berg2015}.
In the kind of interferometry that we are considering, the experiment consists of the following four steps\cite{Mirkhalaf2018,Smerzi2018}:
\\ \\
\indent (1) \textit{State preparation:} In the first step the state, described by a density matrix $\rho$, is prepared.  This step often involves condensing particles into a single wave function and performing entangling operations to allow sensing at higher accuracy.

\textit{(2) Phase encoding:} The unitary evolution of the interferometer encodes a phase onto the state prepared in the first step: $\rho \rightarrow \rho_{\psi}= U^{\dagger}_{\psi,\Omega}\rho U_{\psi,\Omega}$. The Hamiltonian of this unitary evolution is proportional to the parameter to be measured, such as the magnetic field strength. $\psi$ is the phase encoded, and $\Omega$ represents the additional parameters of the unitary transform.

\textit{(3) Read-out:}  An additional unitary evolution $U_{r}$ is applied to the state to prepare for an effective measurement of an observable $R$.

\textit{(4) Projective measurement:} A destructive measurement of an observable $X$ is modelled as a projection onto the eigenvector $\left|x\right>$ with measurement value $x$: $\left<x\right|U_{r}^{\dagger} U_{\psi,\Omega}^{\dagger}\rho U_{\psi,\Omega}U_{r}\left|x\right>$.  Repeating this measurement multiple times produces a distribution:
\begin{equation}
p(r,\psi,\Omega)=\left<r\right|U^{\dagger}_{\psi,\Omega}\rho U_{\psi,\Omega}\left|r\right>
\label{eq:prob4}
\end{equation}
with $\left|r\right>=U_{r}\left|x\right>$.

For simple set-ups, the expectation value of $R$ is directly proportional to the phase encoded and Hamiltonian parameter being estimated.
In this paper, instead of using the last 3 steps to estimate the phase, they are used to verify the indefiniteness of some property of the initial state $\rho$. 

\subsection{Phase encoding, read-out, projective measurement}
A simple form of interferometry involves two quantization modes that can interfere.
These modes can be external kinetic modes in which bosons move in two different guides, or the modes can be identified with the {two} sites of a double well potential\cite{Berg2015,Rosi2015,Debs2011,Albiez2005a, Gross2006, Julia-Diaz2012, Baumann2011,Kovachy2015}.
These modes could also be associated with two different internal states of the boson particles (e.g. hyperfine states of the bosonic atoms\cite{Muessel2014,Strobel2014,Zibold2010,Riedel2010} that can be coupled by lasers).
A highly successful approximation\cite{Raghavan1999,Micheli2003, Leggett2001,Gati2007} assumes that the bosons only occupy these two modes.
This limits the Hilbert space to that spanned by the Fock-states of the two modes: $\left| m_1, m_2\right>$, where $m_1$ and $m_2$ are the number of bosons in the first and second modes.
Counting the particles in the two modes constitutes the projective measurement of step 4: $\left|x\right>=\left| m_1, m_2\right>$

A single particle in two modes has a two dimensional Hilbert space and is described by a spin half operator, ${\bf J}={\bf \sigma}/2$.
The single particle observable in 2 modes, for a system with $N$ particles are described by linear combination of $SU(2)$ generators of a $N/2$ spin, ${\bf J}=\sum_{i=1}^N{\bf \sigma_i}/2$:
\begin{equation}
    J(\theta,\phi) = J_z\cos\theta + 
    J_x\sin\theta \cos\phi + J_y\sin\theta \cos\phi 
\end{equation}
where these Cartesian components, $J_z,J_x$ and $J_y$, satisfy the standard commutation relations: $[J_i,J_j]=i\epsilon_{i,j,k}J_k$.
By mapping the sum, $m_1+m_2=2j$, and difference, $m_1-m_2=2j_z$, onto the magnitude and z-projection of a collective spin, one can connect the Fock representation with this well-known $SU(2)$ algebra for describing rotations.
The particle number difference is then mapped to $J_z$ and tunnelling between the two modes is described by $J_x$(more generally $J(\pi/2,\phi)$).

For internal modes, a Hamiltonian $J_z$ can be created by applying a magnetic field to split the hyperfine states and a Hamiltonian $J_x$ can be created by applying a Rabi-coupling laser field.
For external kinetic modes, these Hamiltonians are controlled by shaping the external potential. 

The phase encoding and read-out operations, $U_{\psi,\Omega}$ and $U_r$ discussed in this paper, are all linear single-particle operations:
\begin{equation}
    U(\alpha,\theta,\phi) = e^{-i\alpha J(\theta,\phi)}
    \label{eq:rot}
\end{equation}
where $U_r=U(T_r\epsilon_r/\hbar,\theta_r,\phi_r)$ and $U_{\psi,\Omega}=U(\psi,\theta_{\Omega},\phi_{\Omega})$ and $\epsilon$ is the energy scale of the Hamiltonian.
The collective spin picture maps a projective two-mode number difference measurement, $m_1-m_2$, to a measurement of the $J_z$ observable.
For single-particle read-out, the combined steps 3 and 4 becomes equivalent to an effective measurement on the spin in a new direction:$J_z\rightarrow J_{z'}= U_r^{\dagger}J_z U_r$.
For example, a read-out rotation around the x-axis ($T_r=\frac{\pi\hbar}{2\epsilon_r}$, $\theta_r=\pi/2$, $\phi_r=0$. \textit{i.e.} $J(\theta_r,\phi_r)=J_x$) produces an effective measurement of $J_y$.

\subsection{State Preparation: Cat States and Mixed Double-Peak States}

\begin{figure}
    \includegraphics[width=0.5\textwidth]{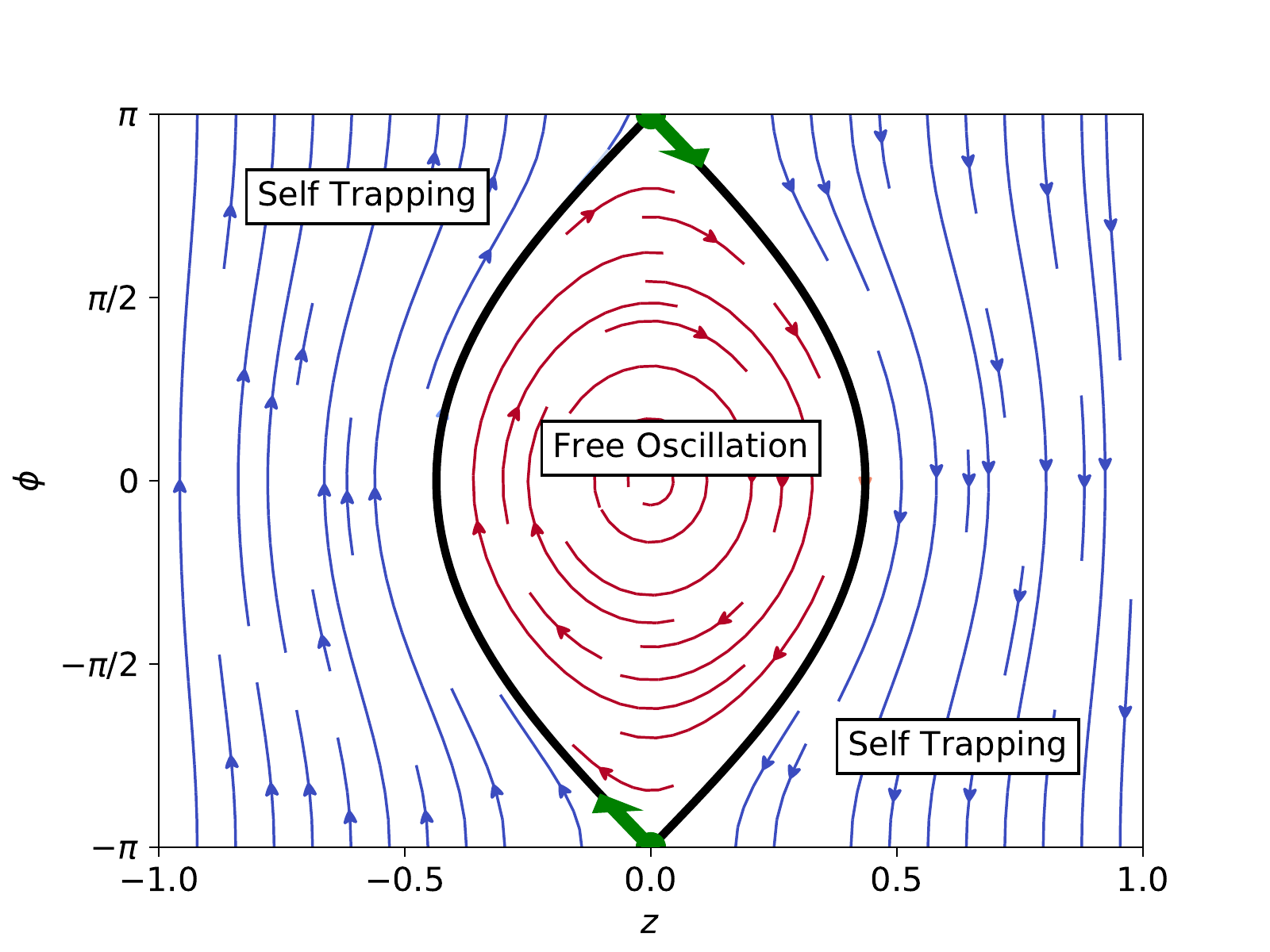}
    \caption{(Color online) Classical trajectories. The separatrix is shown in black (bold) and separates the circular free-oscillation trajectories from the self-trapping ones. The green dots mark the fixed point and the green arrows mark the unstable directions.}
    \label{fig:classicalFlow}
\end{figure}

In this paper, the state prepared in the first step of interferometry is the Gordon and Savage cat state\cite{Gordon1999} or a mixed state with a similar distinction between dead and alive states.
In this section, we describe these states and how they can be prepared.
In the next section we describe how the last 3 steps can be used to verify indefinite properties of this state.

In bosonic interferometry, state preparation begins with condensation into the ground state of some Hamiltonian $\epsilon_\tau J(\theta,\phi)$.
In this paper, we will describe partial condensation using a thermal state:
\begin{equation}
    \rho(\beta,z,\phi) = e^{\beta \epsilon_{\tau} J(\cos^{-1}(z),\phi)},
    \label{eq:thermal}
\end{equation}
where we have introduced the scaled difference: $z=j_z/j=\left[m_1-m_2\right]/N=\cos\left(\theta\right)$.
States of this form have been produced for kinetic modes for the Hamiltonian $J_x$ by Gross \textit{et al.}\cite{Gross2006}, and thermal states of any other Hamiltonian of the form $J(\cos^{-1}(z),\phi)$, can be produced by rotations of the form in Eq. \ref{eq:rot}.
For this paper, we will focus on the states $\rho(\beta,0=z_c(\pi),\pi)$ and $\rho(\beta,\left|z_c(0)\right|,0)$ which we refer to as the $\pi$ and $0$ state at temperature $\beta^{-1}$.
The critical imbalance, $z_c(\phi)$, is given by the solid black line in Fig.~\ref{fig:classicalFlow}.

Following (partial) condensation, cat states can be prepared by the method mentioned above by Gordon and Savage.
We will describe this method using the explanation provided by Micheli \textit{et al.}\cite{Micheli2003}.
There they explain how the twist-and-turn\cite{Smerzi2018} Hamiltonian:
\begin{equation}
    H = t J_x + \frac{U}{2}J_z^2  \label{eq:ham}
\end{equation}
produces cat states via a semi classical analysis.
The classical analysis assumes a set of variational states which are the ground state of the Hamiltonian $J(\cos^{-1}(z),\phi)$.
The classical equations of motion describe the dynamics of imbalance of particles between the two modes (projection onto the axis), $z$ and its conjugate variable $\phi$.

The classical equations of motion have been solved analytically\cite{Raghavan1999} and have two fixed points for all parameters, $t$ and $U$.
Dynamical creation of a cat state takes place at larger coupling strength ($U>4t/N$), where one of the classical fixed points is unstable.
The classical trajectories for $2J=N=200$, $U=0.1$ and $t=1$ are shown in Fig.~\ref{fig:classicalFlow} and demonstrates a critical line, $\pm z_c(\phi)$, separating two distinct dynamical behaviours.
Along one set of trajectories the effective spin rotates around the x-axis so that the variation of the azimuthal ($\phi$)-angle is confined to a finite
interval.
These trajectories, confined to the middle region in Fig.~\ref{fig:classicalFlow}, correspond to the {Josephson oscillations\cite{zapata1998, Raghavan1999,Albiez2005a}} observed in condensed matter Josephson junctions and we refer to them as ‘free oscillations’.
Along another set of trajectories, the spin rotates around the z-axis so that the $\phi$-variable increases indefinitely.
Along the latter type of trajectories the particle imbalance ($z$) does not change sign and the corresponding dynamics is known as ‘self-trapping’ dynamics.
The phase-space ($\phi,z$)- regions of the two types of trajectories are separated by a critical line, $z_{c}(\phi)$, called the separatrix, indicated by the thick black line in Fig.(1). 
This line is the classical trajectory of both the $\pi$ and $0$ states in the classical analysis.
The $\pi$ state starts on the unstable fixed point, while the $0$ starts at $(z_c(0),\phi=0)$.
All numerical calculations presented in this paper have been carried out for the parameters used in Fig.~\ref{fig:classicalFlow}: $2J=N=200$, $U=0.1$ and $t=1$.

Focusing on pure states ($\beta^{-1}=0$), the first quantum approximation in a semi-classical analysis treats the initial pure state as a finite width Gaussian probability distribution. 
In the classical dynamics, the paths of the free oscillation and self-trapping trajectories diverge near the unstable fixed point($\phi=\pi,z=0$).
In the quantum mechanical evolution of the $\pi$ and $0$ states, the trajectories of the Wigner-distribution amplitudes part ways near the same phase space coordinate.
After a time interval during which the z-coordinates of the classically evolving systems on either side of the separatrix have separated maximally (a time $T_{\pi}=log(8N) \hbar/[N U]$ for the $\pi$-state and $1.4 T_{\pi}$ for the $0$-state), the quantum state evolves into a superposition of two macroscopically separated (specified below) states.
The corresponding self-trapping and free oscillation components are the dead and alive components of the cat state.
In the case of the pure state described above, we a priori know that the vital status is indefinite.

For the case in which the initial state is at a high temperature, the classical dynamics are the same, but uncertainty in the evolved probability distribution reflects our lack of knowledge about the classical phase-space position as opposed to the indefiniteness of the quantum state.
As we show below, the measure that we propose indicates that the thermal states are definite.

We numerically compute both the thermal and pure states using exact diagonalization of Eq. \ref{eq:ham} followed by a time evolution of the states $\rho(\beta,z,\phi)$.
The probability distributions for the $J_z$ observable are shown in Fig. \ref{fig:probs} for the pure states and Fig.~\ref{fig:probsT} for the high temperature states.
Both the pure states and the thermal states demonstrate a double peak suggestive of a dead and alive labelling.
We make this labelling precise for a pure state $\left| k\right>$ by decomposing it into a dead and alive state$\left| k \right> = (\left|alive\right> + \left|dead\right>)/\sqrt{2}$:
\begin{eqnarray}
    \left|alive\right> = \frac{1}{\sqrt{N_L}}\sum_{j_z}\left|j_z\right>\left<j_z|k\right>\Theta(\left<J_z\right>-j_z) \\ \nonumber
    \left|dead\right> =  \frac{1}{\sqrt{N_R}}\sum_{j_z}\left|j_z\right>\left<j_z|k\right>\Theta(j_z-\left<J_z\right>)
\end{eqnarray}
Where $N_R$ and $N_L$, are defined so the alive and dead states are properly normalized. 
While this decomposition is always possible, it only make sense to call the pure state $\left|k\right>$ a cat state if the dead and alive states are macroscopically distinct. 
In other words, the extensive difference $\Lambda(A)=\left<A\right>_{alive}-\left<A\right>_{dead}$ should scale with the number of particles.
For the decomposition above, the extensive difference for the observable $J_z$ can then be computed by:
\begin{equation}
    \Lambda(J_z) = \sum_{j_z} (P_L(j_z)-P_R(j_z))j_z
    \label{eq:ed}
\end{equation}
where $P_L(j_z)$ are the re-normalized distributions corresponding to the dead and alive states:
\begin{eqnarray}
    P_L(j_z) = \frac{1}{N_L}P(j_z)\Theta(\left<J_z\right>-j_z) \\ \nonumber
    P_R(j_z) = \frac{1}{N_R}P(j_z)\Theta(j_z-\left<J_z\right>)
\end{eqnarray}
{ where $\Theta(x)$ is the Heaviside-step function which is 1 for $x>0$ and 0 for $x<0$.}

\begin{figure}
       \includegraphics[width=0.4\textwidth]{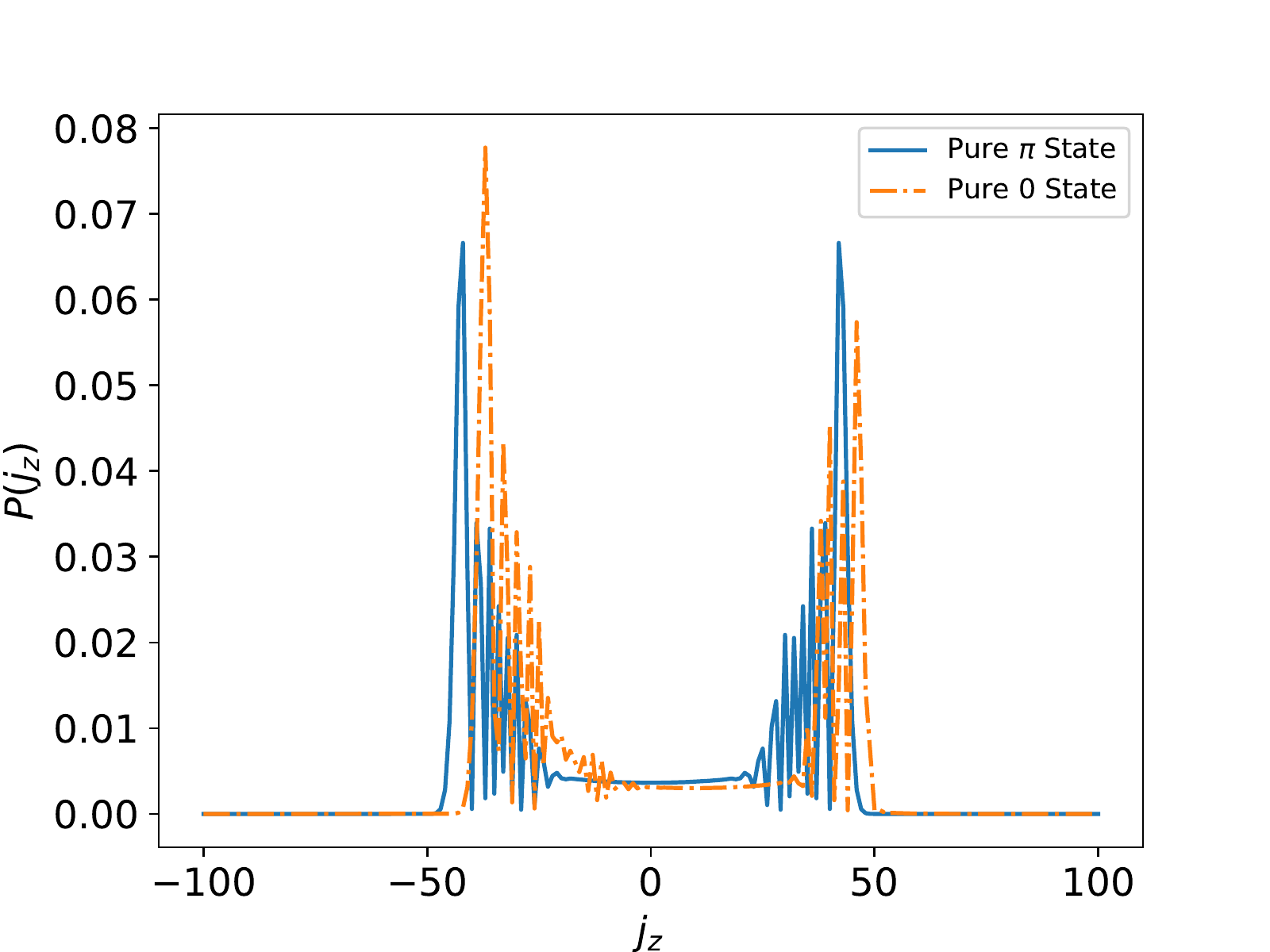}
       \caption{Distributions $P(j_z)$ for the state evolved from the $\pi$ and $0$ coherent states ($0$ temperature) for a time $T_\pi$ and $1.4T_\pi$ respectively. These states were computed for $N=200$ (spin with size $J=100$), and the x-axis, $j_z$, are the eigenvalues of the observable $J_z$.}
    \label{fig:probs}
    \centering
       \includegraphics[width=0.4\textwidth]{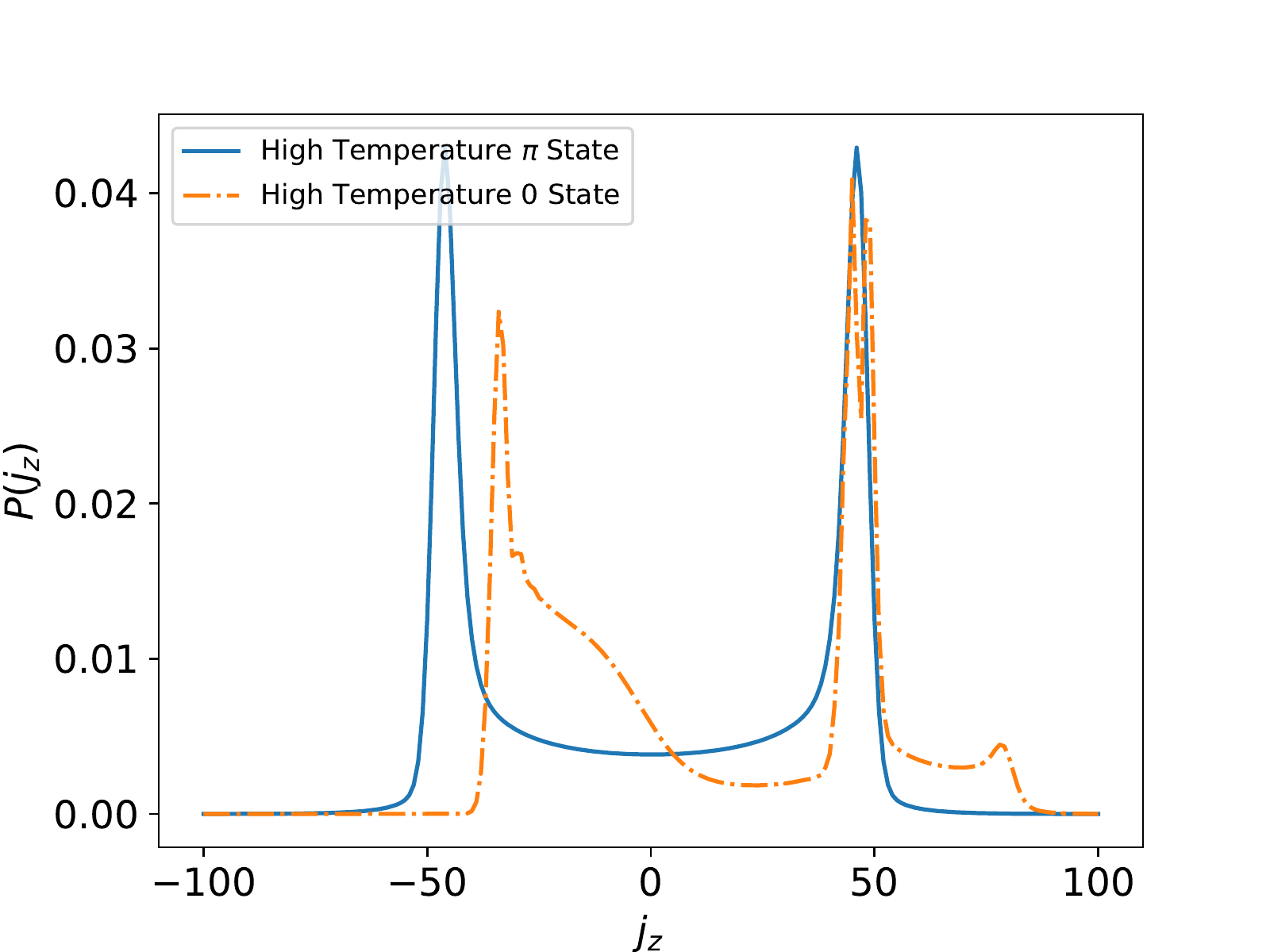}
       \caption{Distributions $P(j_z)$ for the state evolved from the $\pi$ and $0$ states at temperature $10\epsilon_\tau$ for the same time intervals as in Fig.~\ref{fig:probs}. As in Fig.~\ref{fig:probs}, these states were computed for $N=200$ (spin with size $J=100$).}
    \label{fig:probsT}
\end{figure}

Thus, any double peak distribution where the peaks are macroscopically separated will have an extensive difference scaling with the number of particles.
This is true for the pure states in Fig.~\ref{fig:probs}, where the extensive difference is $65 \approx 200/3 = N/3$.
Eq.~\ref{eq:ed} can also be applied to the mixed states in Fig.~\ref{fig:probsT} and gives a similar extensive difference $\approx N/3$.
Experimentally, counting statistics provide the distributions $P(j_z)$ and Eq.~\ref{eq:ed} can be used to determine if the observed state can be meaningfully separated into macroscopically distinct dead and alive cats.
The next section describes how to determine the vital status of the cat.

\section{Interferometer Sensitivity and Indefiniteness}
\label{qfisec}
To quantify how indefinite the vital status of the cat is, we use the interferometer sensitivity quantified by the quantum and classical Fisher information (QFI and CFI).
In this section, we introduce the interferometer sensitivity, convex roof of the variance, and explain how an experiment can quantify the indefiniteness and obtain information about possible pure states which make up the density matrix. 

{In the interferometry experiment discussed above, a phase $\psi=\frac{\epsilon \Delta t }{\hbar}$, is encoded on to a state via time evolution by a  Hamiltonian, $H_{\Omega}= \epsilon J\left(\theta_{\Omega},\Phi_{\Omega}\right)$, for a time $\Delta t$.
The sensitivity to the phase $\psi$ is given by the CFI}:
\begin{eqnarray}
\label{eq:CFI}
    F_{c}(R,\rho_{\psi},\Omega) =\sum_{r}p(r,\psi,\Omega)[\partial_{\psi}log(p(r,\psi,\Omega)))]^2
\end{eqnarray}
where $p(r,\psi,\Omega)=\left<r\right|U^{\dagger}_{\psi,\Omega}\rho U_{\psi,\Omega}\left|r\right>$(Eq.~\ref{eq:prob4}).
The primary use of the CFI, $F_c$, is that its value provides an upper bound on the estimated phase $\psi$ via the Cramer-Rao bound\cite{cramer1945,rao1945}:
\begin{equation}
    \Delta \psi \geq \frac{2}{\sqrt{ F_{c}(R,\rho_{\psi},\Omega)}}
\end{equation}

The CFI, $ F_{c}(R,\rho_{\psi=0},\Omega)$, can be measured in experiments\cite{Strobel2014}: repeating the four-step process to obtain measurements of $p(r,0+\delta,\Omega)$ for a range of small $\delta$ allows the construction of the derivative with respect to $\psi$ evaluated at $\psi=0$ and a direct use of Eq.~\ref{eq:CFI}. Other methods exist to get more accurate values \cite{Mirkhalaf2018,Garttner2018,Frowis2016}.

A pure state with larger uncertainty, $\Delta J(\theta_{\Omega},\phi_{\Omega})$, (implying, since the state is pure, a larger indefiniteness in the observable $J(\theta_{\Omega},\phi_{\Omega})$), responds on a faster time scale, $\omega^{-1} = \hbar( \epsilon \Delta J(\theta_{\Omega},\phi_{\Omega}))^{-1}$ and may have a larger CFI.
Whether or not the CFI is larger depends on the observable $R$ in step 3 of the 4 step process: the dependence on the phase ($\psi$) cancels out if the  $\{\left| r \right>\}$ basis consists of eigenstates of $J(\theta_{\Omega},\phi_{\Omega})$.
To characterize the sensitivity of the quantum state, independent of the choice of the observable $R$, one must optimize over all Hermitian operators $R$. The result of this optimization procedure is the QFI\cite{caves1994,Toth2014,Frowis2012a}:
\begin{equation}
    F_{q}(\rho_{\psi},\Omega)=\max_{R}F_{c}\left(R,\rho_{\psi},\Omega\right)
    \label{eq:qfilim}
\end{equation}
Since the Cramer-Rao uncertainty bound on $\Delta \psi$ of Eq.(10) is valid for every choice of the measurement observable, $R$, the tightest bound on $\Delta \psi$ is  obtainable from the QFI:
\begin{equation}
    \Delta \psi = \frac{\epsilon \Delta t}{\hbar}  \geq \frac{2}{\sqrt{F_{q}\left(\rho_{\psi},\Omega\right)}}.
\end{equation}
For a pure state system, $\rho = \left|k\right>\left<k\right|$, it was shown\cite{caves1994} that 
\begin{equation}
    F_{q}\left( \rho_{\psi}=\left|k\right>\left<k\right|,\Omega\right) =  4 \left<k\right| \left( \Delta J \left(\theta_{\Omega},
\Phi_{\Omega}\right) \right)^{2} \left|k\right> 
\end{equation}
where
\begin{equation}
    \left<k\right|(\Delta J)^2\left|k\right>=\left<k\right| J^2\left|k\right>-\left<k\right| J\left|k\right>^2
\end{equation}
With $\Delta \psi = \left( \epsilon \Delta t \right)/\hbar$, and
{$\epsilon \Delta J = \Delta H_{\Omega}$,}
the pure state Cramer-Rao bound on the phase can be written as
{\begin{equation}
    \Delta t \sqrt{\left<k\right| \left(\Delta H_{\Omega}\right)^{2} \left|k\right>} \geq \hbar  
\end{equation}}
in agreement with the Heisenberg energy-time inequality.

Here we have chosen to consider the sensitivity of the state $\rho_{\psi=0}$ with $0$ phase encoded because we are interested in properties of the state evolved after the first-step, not a different state with phase encoded onto it.

Since the state is pure, we know that any uncertainty in an observed property of the state directly corresponds to a quantum phenomenon of indefinite properties.
For a mixed state ensemble, it is not immediately clear that the QFI generalizes the statistical variance as a quantification of indefiniteness.
To address this, S. Yu\cite{Yu2013} and Toth \textit{et al.}\cite{Toth2013} proved the following illuminating expression for the QFI:
\begin{equation}
    F_{q}(\rho_{\psi=0},\Omega)  =\min_e \sum_{k}P_{k_e}F_q(\left|k_e\right>\left<k_e\right|,\Omega)
    \label{eq:QFI}
\end{equation}
where the optimization over $e$ is over all decomposition of a density matrix, $\rho_{\psi=0}$, into an ensemble of pure states $\rho_{\psi=0}=\sum_{k}P_{k_e}\left|k_e\right>\left<k_e\right|$, where the  $\{\left|k\right>\}$-states of {this} decomposition are not necessarily orthogonal. This decomposition is not unique because in the vector space of density matrices, the set of all pure state density matrices form an over-complete basis. Thus, $e$ represents one of these non-unique decompositions and  $k_e$ labels the pure states which make up that decomposition.
The right hand side of Eq.~\ref{eq:QFI} is known as the convex-roof of the variance\cite{bengtsson_zyczkowski_2006,Yu2013,Toth2013}.

Cast as a generalization of the concept of statistical variance, the QFI, $F_{q}$, can be seen to provide a valid measure of indefiniteness.  Indeed, the minimization in the space of density matrices implies that a portion {of} the sum, $\sum_{k} P_{k_{e}} \left|k_{e}\right>\left<k_{e}\right|$,  of significant $P_{k_{e}}$-weight involves pure states, $\left|k_{e}\right>$,  with a statistical variance  $\left< k_{e} \right| \left( \Delta J ( \theta_{\Omega}, \Phi_{\Omega}) \right)^{2} \left| k_{e} \right>$ that is  comparable to the convex uncertainty:
\begin{equation}
    \label{eq:convexDelta}
    \Delta_q J(\theta_{\Omega},\phi_{\Omega})={\frac{1}{2}}\sqrt{F_{q}(\rho_{\psi=0},\Omega)}
\end{equation}
This implies that, if we were given full knowledge of the universe, and were able to sort the results based on which pure state, $\left|k_e\right>$, was produced by the experimental apparatus, the majority of the distributions, $P_{k_e}(j)=\left|\left<j(\theta_{\Omega},\phi_{\Omega})|k_e\right>\right|^2$, would have a statistical uncertainty larger than $\Delta_q J(\theta_{\Omega},\phi_{\Omega})$. Since this uncertainty can not be reduced by obtaining more information, it must be due to the indefiniteness of the observed property $J(\theta_{\Omega},\phi_{\Omega})$.

Thus a measurement of large sensitivity, $F_{q}(\rho_{\psi=0},\Omega)$, implies a large indefiniteness of the phase encoding Hamiltonian $\epsilon J(\theta_{\Omega},\phi_{\Omega})$ in the initial state $\rho$.
This was pointed out by Fr{\"o}wis and D{\"u}r\cite{Frowis2012a}, and was used to construct a measure of indefiniteness `$N_{eff}$' (defined below in Eq.~\ref{eq:neff}) by the way $\max_\Omega F_{q}(\rho_{\psi=0},\Omega)$ scales with the number of particles.
{ As discussed in the Appendix \ref{apdx:flasePositive}, this optimization over observables $\Omega$ can lead to misleading results when considering the indefiniteness associated with the superposition of two macroscopically distinct states.
Instead,} we use the extensive difference for an observable $J_{\Omega}$ as a measure of the size of the cat, and we introduce the comparison of the convex uncertainty with the statistical uncertainty
\begin{eqnarray}
    r_q(\Omega) = \frac{\Delta_q\left( J_{\Omega} \right)}{\Delta_{s}\left( J_{\Omega} \right)} = \frac{\Delta_q\left( J_{\Omega} \right)}{\sqrt{Tr[J_{\Omega}^2\rho]-Tr[J_{\Omega}\rho]^2}}
\end{eqnarray}
as a measure of the quality of indefiniteness.  Since the statistical uncertainty is always greater than the convex uncertainty, $r(\Omega)$ ranges from 0 to 1.
When $r_q(\Omega)$ is 1, any observed statistical uncertainty is due to indefiniteness, while for smaller $r_q$, only a fraction of the uncertainty is due to indefiniteness. The statistical uncertainty can be obtained as part of the same interferometry experiment{: if the interferometric procedure is repeated with $\psi=0$, and with the effective observable  as $R=J_\Omega$,} the statistical uncertainty follows from counting the $p_r$-distributions obtained after these steps.

We use $r(\Omega)$ and $\Lambda(J_\Omega)$ because, with the additional knowledge of a double peak distribution in the observable $J_\Omega$, qualitative arguments can be made about the amplitudes of pure states which could make up a representative density matrix ensemble, $e=\{P_{k_e},\left|k_e\right>\}$.
If the convex and statistical uncertainties are approximately equal to each other, $\Delta_{s}/\Delta_{q} \approx 1$, we know the density-matrix ensemble is, on average, composed of pure states with uncertainty similar to that of the observed statistical distribution.
In addition, since different pure states in the ensemble can not destructively interfere with each other,
    we know that the pure states in the density matrix have small amplitude for the basis states that have small probability of occurrence in the full statistical ensemble.
Thus with the additional observation of a double peak, we can conclude that any representation of the density matrix is mostly composed of cat states with extensive difference similar to the observed one.

What can be said when $r_q$ is not very close to 1, but still significant (\textit{e.g.} $r_q>0.1$)?
To answer this question we introduce the product $\Lambda r_q$ as the ``reduced extensive difference", where the extensive difference $\Lambda$ is given by Eq.~\ref{eq:ed} .
As long as the individual peaks have narrow width (similar to the pure cat states, see Fig.~\ref{fig:probs}) and the reduced extensive difference is significantly larger than the peak width, we can again qualitatively argue that there exists pure states in the density matrix ensemble with extensive difference similar to that of the observed extensive difference $\Lambda$.
If the reduced extensive difference is significantly larger than the width of the peak, there must exist pure states, $\left|k_e\right>$ with variance significantly larger than the width of the peaks and are realized with significant probability $P_{k_e}$.
Since the peaks are narrow and there is very low probability between the two peaks, the only form these states can take is one with double peak amplitudes similar to the observed distribution.
Thus, we know the density matrix contains a significant off diagonal contribution $\left<m|k_e\right>\left<k_e|m'\right>P_{k_e}$ for $|m-m'|\approx \Lambda(J_z)$, despite {an imperfect quality of indefiniteness, $r_1<1$.}
This makes a connection with the work done by Opanchuk \textit{et al.}\cite{Opanchuk2016}, who put bounds on $\left<m\right|\rho\left|m'\right>$ using multi-particle correlation functions.

\section{Results: Detection Of Indefiniteness via Interferometer Sensitivity}

\begin{figure}
    \includegraphics[width=0.45\textwidth]{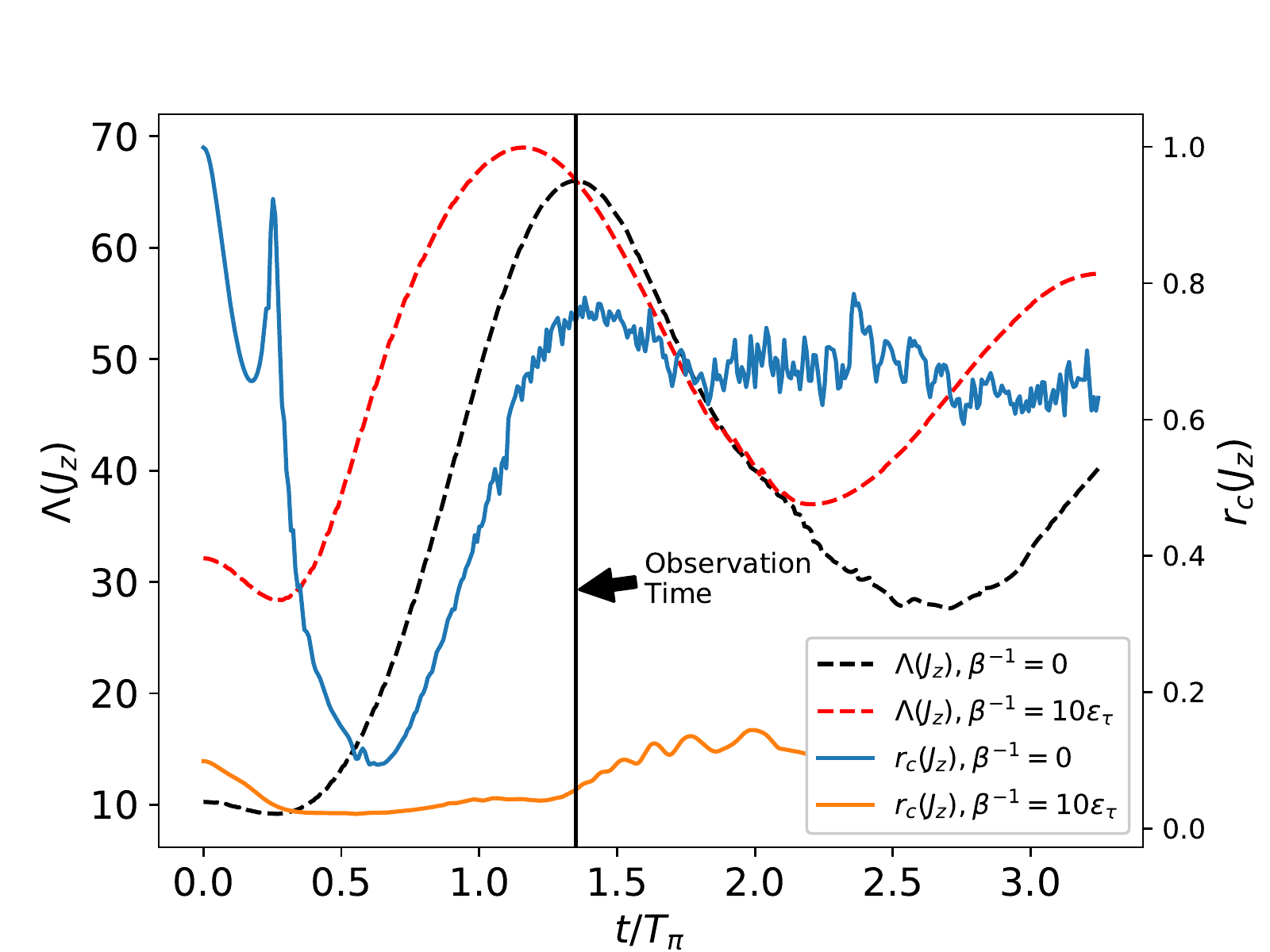}
    \caption{In this plot, the extensive difference (dashed lines, left axis) and quality of indefiniteness (solid lines, right axis) are plotted versus the time spent during the non linear evolution in step 1 of the interferometric process. The scale of the extensive difference is set by the number of particles $N=200$ (spin of size $J=100$), in the evolving state. These quantities are computed for the $0$ state at temperature $0$ and $10\epsilon_\tau$.  An experimenter, testing these cat states shown, would compute the statistical variance and extensive difference from the double peak distributions shown in Fig.~\ref{fig:probs}. They would perform the interferometry process discussed above to compute the bound on $\Delta_q$ via the CFI.}
    \label{fig:lrvTIME}
\end{figure}

In this section, we describe how an experiment would observe the measures discussed in the previous section and what they would observe for the Gordon-Savage cat state and the mixed states discussed in Section I.
The simplest step in such an experiment requires measuring the probability distributions in Fig.~\ref{fig:probs}.
This requires the state preparation described in Section I, followed by the projective measurement without any phase encoding or read-out. 
Repeating this reconstructs the distributions for the observable $J_z$.
The extensive difference can then be computed by Eq.~\ref{eq:ed}.

For completeness, we have plotted (Fig.~\ref{fig:lrvTIME}) the dependence of the extensive difference for the $0$ thermal state at $\beta^{-1}=0$ and $\beta^{-1}=10\epsilon_{\tau}$ versus the time spent during the non linear evolution that creates the cat.
The extensive difference reaches a maximum at a time $1.4*T_{\pi}=1.4\log(8N)\hbar /NU$ ($\beta^{-1}=0$) and $1.1T_{\pi}$($\beta^{-1}=10\epsilon_{\tau}$).
The probability distributions $P(j_z)$ at these times are shown in Fig.~\ref{fig:probs}.
An experimenter interested in a specific cat does not need to measure the extensive difference at all times.
Rather, they can do measurements at time $1.4T_{\pi}$ for the $0$ states or $T_\pi$ for the $\pi$ states\cite{Micheli2003}.

This calculation shows that, for the parameters considered ($N=200$, $U=0.1t$), the extensive difference of the $0$ state is expected to peak at $1.4 T_\pi$, and thus suggests $1.4T_\pi$ as a good time to end state creation (step 1) and begin the statistical and interferometric measurements (steps 2-4).
{Measuring the distribution $P(j_z)$ at this time,} they will find an extensive difference of $65 \approx 200/3=N/3$ particles (Fig.~\ref{fig:lrvTIME}).
{For an experiment performed for a fixed particle number, the difference in expectation values between the dead and alive cats (i.e. the extensive difference $\Lambda$) would be on the same order of magnitude as the number of particles.
An experiment could then be repeated for different number of particles, and would find the extensive difference scales with $N$\footnote{We directly confirmed this numerically by computing the $\pi$ and $0$ states for $N=200\dots800$ and found a linear scaling of the extensive difference with the number of particles as $N/3.1$. The semi-classical approach also predicts a linear in $N$ scaling\cite{Micheli2003}}, suggesting that if the trends continue, a macroscopic number of particles would yield a macroscopic cat state.}

The second step is to verify the indefiniteness of the cat's vital status.
Here, one should compare the statistical and convex uncertainty of the observable $J_z$, because this was the observable which demonstrated the macroscopic difference ($\Lambda (J_z)$) between the dead and alive cats.
The statistical uncertainty can be computed directly from the distributions in Fig.~\ref{fig:probs}.
The convex uncertainty (computed from the QFI using Eq.~\ref{eq:convexDelta}) for $J_z$ is bounded by measuring the sensitivity (CFI) of a probability distribution for some observable $J_r$ to a phase encoding operation $J_{\theta_{\Omega}, \phi_{\Omega}}=J_z$.
The single-particle observables that provide the best bounds will be the ones that respond  most to rotations around the z-axis: any spin pointing in the x-y plane.
We use $J_r=J_y$, since rotations around the x-axis are easily implemented, as described in Section I.
Experimentally, the interferometric process is repeated with $J_r=J_y$, $J_\Omega=J_z$ and $T_r=\frac{\pi\hbar}{2\epsilon_r}$ for multiple small $\psi\epsilon_\Omega=0+\delta$, such that the distribution $p(j_y,0,J_z)$ and its derivative can be computed and used in the expression for the CFI(Eq.~\ref{eq:CFI}).
With a measurement of the CFI, one can bound the convex uncertainty and quality of indefiniteness via Eq.~\ref{eq:qfilim}:
\begin{eqnarray}
    r_q(J_z) = \frac{\Delta_q(J_z)}{\Delta_s(J_z)}>r_c(J_z)=\frac{{\frac{1}{2}}\sqrt{F_c(R=J_y,\rho,J_z)}}{\Delta_s(J_z)}
\end{eqnarray}

Using the statistical distribution for $\Delta_s$, and:
\begin{eqnarray}
\label{eq:CFIcalc}
    F_{c}(R,\rho_{\psi},\Omega) = -\sum_{r}\frac{1}{ p(r,\psi)}\left<r\right|[\rho_{\psi},\epsilon J(\theta_{\Omega},\phi_{\Omega})]\left|r\right>^2 \nonumber
\end{eqnarray}
for the CFI\footnote{This expression can be obtained by expanding the unitaries in Eq.~\ref{eq:prob4} around perturbations about $\psi$, and substituting into Eq~\ref{eq:CFI}.}, we numerically compute (and plot in Fig.~\ref{fig:lrvTIME}) $r_c(J_z)$ for the $0$ thermal state at $\beta^{-1}=0$ and $\beta^{-1}=10\epsilon_{\tau}$ versus the time spent during the non linear evolution which creates the cat.
For the pure state ($\beta^{-1}=0$), $r_q=1$, and $r_c<1$ reflects the imperfect bound the choice of the observable $R=J_y$ puts on the QFI.
For the cat state produced after a non linear evolution for $t=1.4T_\pi$, the quality of indefiniteness measured by an experiment is about 0.75(see  Fig.~\ref{fig:lrvTIME}).
Furthermore, the reduced extensive difference, $\Lambda r_c\approx N/4=50$ is significantly larger than the width of the peaks (approximately $N/20=10$). Thus, in good faith, an experimenter can believe that the density-matrix ensemble which they are observing is mostly composed of pure states with double peak amplitudes.
Furthermore, {since $r_q=1$,} one can expect to be able to account for 100\% of the quantum variance by using a more optimal observable $R$ \cite{Mirkhalaf2018,Garttner2018,Frowis2016,Oberthaler2016}.

For the state evolved (at $t=1.1T_\pi$) from the high temperature distribution ($\beta^{-1}=10\epsilon_{\tau}$), the quality of indefiniteness is 5\% and the reduced extensive difference is $\Lambda r_q=3=O(1)$.
This is smaller than width of the peak.
We must therefore conclude that there is no indefiniteness and that the cat is not dead and alive at the same time.
Even with the ideal bound (see Fig.~\ref{fig:lrVtemp}), the reduced extensive difference is still on the same size as the peak width ($\Lambda r_q=10$).

\begin{figure}
    \includegraphics[width=0.5\textwidth]{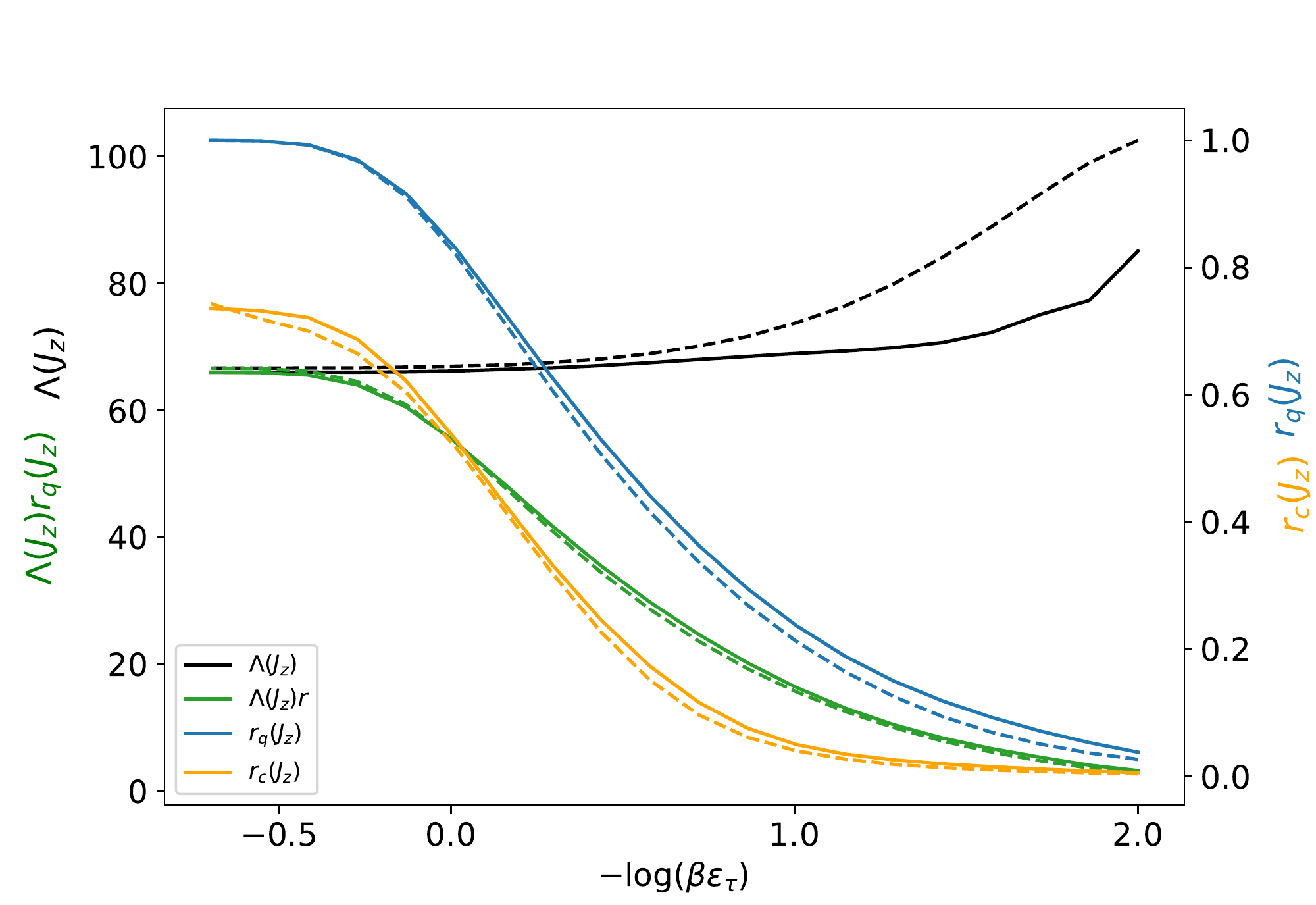}
    \caption{Quality of indefiniteness, $r_q$, its experimental bound $r_c$, the extensive difference $\Lambda(J_z)$ and the reduced extensive difference $\Lambda(J_z)r(J_z)$ are plotted versus the temperature of the initial state.
    The solid lines are for the thermal states at $\pi$, while the dashed lines are for the thermal states at $0$.
    A quantum to classical crossover is shown between temperatures $\epsilon_{\tau}$ and $10\epsilon_\tau$.
    As in Fig.~\ref{fig:lrvTIME}, the scale of the extensive difference is set by the number of particles $N=100$ in the evolved state.
}
    \label{fig:lrVtemp}
\end{figure}

In the remainder of this section, we show how these experiments are capable of detecting the crossover to a classical mixture as the temperature of the initial state is increased.
Fig.~\ref{fig:lrVtemp} demonstrates that the quality of indefiniteness, $r_q$, and its experimental bound $r_c$ decay to 0 as the temperature is increased.
The quantum to classical crossover occurs slowly between $\beta^{-1} = \epsilon_{\tau}$ and $10\epsilon_{\tau}$, where $\epsilon_{\tau}$ sets the energy-scale of the spin Hamiltonian as in Eq.~\ref{eq:rot}.
For $\beta^{-1}<<\epsilon_{\tau}$, the initial state condenses into the pure state and the quantum variance plateaus at its pure state value.
This system is particularly interesting, in that the live and dead cat are still macroscopically different ($\Lambda=O(N)$) even at high temperature.
Since the extensive difference remains constant, the difference between the dead and alive states is still macroscopic, and there are still two macroscopically distinct states which can be labelled dead and alive.
We can then interpret the decay of the quality of indefiniteness to 0 when temperature is increased as a crossover from a cat being dead and alive at the same time to a cat being either dead or alive.

\begin{figure}
    \includegraphics[width=0.5\textwidth]{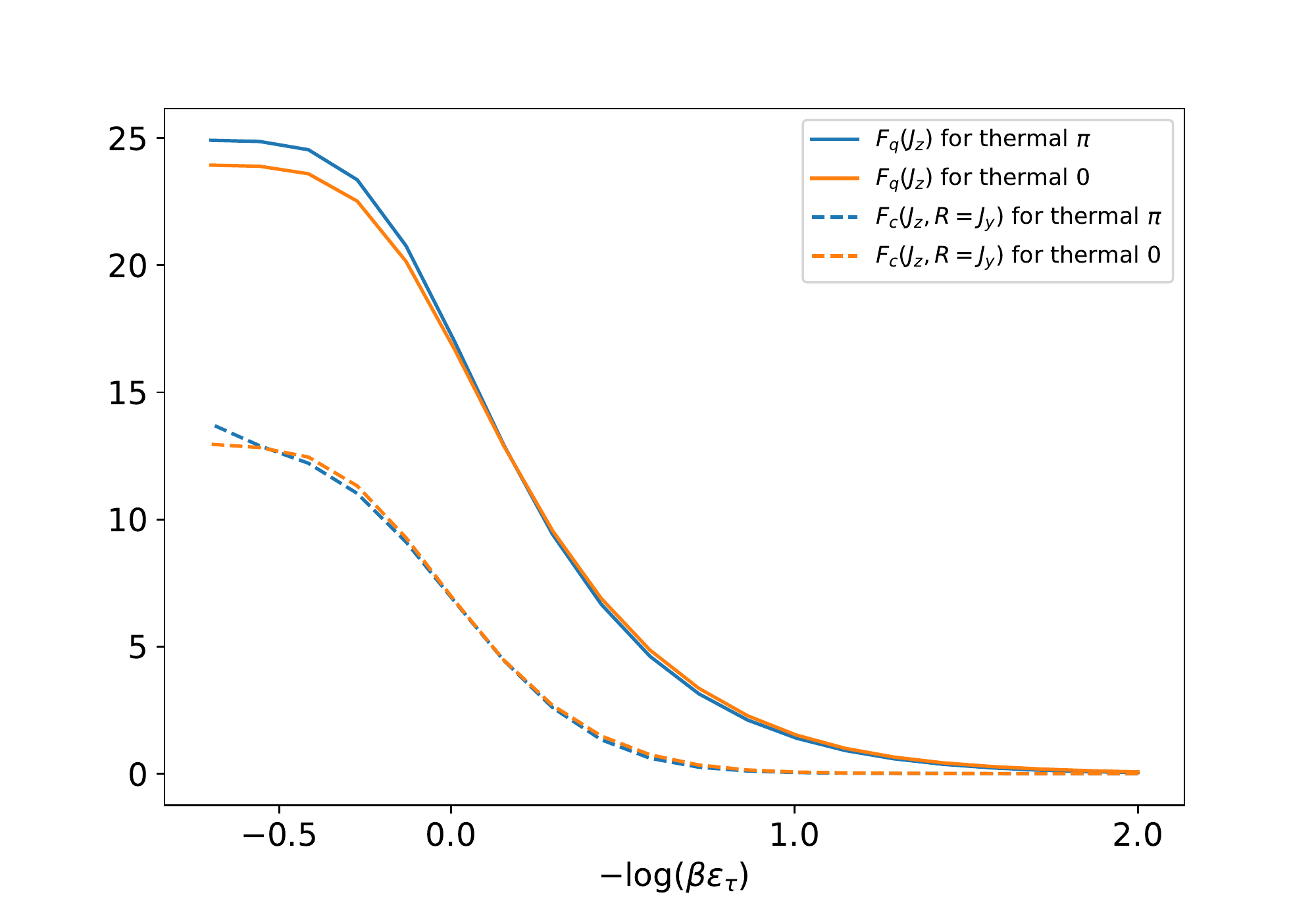}
    \caption{The QFI and its experimental bound for convex uncertainty of $J_z$ versus temperature.  These are lower bounds on the Fr{\"o}wis and D{\"u}r measure (Eq.~\ref{eq:neff}), which also demonstrate the quantum to classical crossover. The y-axis is shown in units of ${\frac{1}{4N}}$.}
    \label{fig:qVar}
\end{figure}

The metric for quantum macroscopicity proposed by Fr{\"o}wis and D{\"u}r\cite{Frowis2012a}, also shows this quantum-classical crossover.
This metric is given by:
\begin{equation}
    N_{eff}=\frac{1}{4N} \max_\Omega F_q(\rho, \Omega)
    \label{eq:neff}
\end{equation}
In addition to other methods\cite{Mirkhalaf2018,Frowis2016,Garttner2018}, this can be experimentally bounded from below using the CFI as done above for $r_q$ (using $r_c$).
The bound provided by $F_q(J_z)$ and its experimental bound $F_c(R=J_y,J_z)$ are plotted in Fig.~\ref{fig:qVar}.
The crossover region is the same for Fr{\"o}wis and D{\"u}r's as for the measures above {($r_q$ and $\Lambda r_q$)} because the statistical variance $\Delta_s$ and extensive difference $\Lambda$ is relatively constant through the crossover region.
Thus, the main difference is the size of the cat each quantify: both are macroscopic in that they are $O(N)$, but the extensive difference is roughly twice as large.
The difference stems from the difference in motivation of the two measures.
The extensive difference attempts to describe the difference between the dead and live cat, while the measure by Fr{\"o}wis and D{\"u}r aim to quantify a relative improvement in sensitivity from unentangled states (such as those in Eq.~\ref{eq:thermal}).
Furthermore, by focusing on the indefiniteness in a specific observable $J_z$, the extensive difference, $\Lambda(J_z)$, and the quality of indefiniteness, $r(J_z)$, provide additional information about the stability of the dead and alive states as the temperature is increased.

Using this lower bound for $N_{eff}$, {a similar conclusion about quality of indefiniteness is reached}, but improving the bound on $N_{eff}$ could lead to different conclusions.
In the appendix we show that the dead and alive states of the cat have macroscopic indefiniteness independent of their superposition.
We can therefore imagine a situation where $N_{eff}$ is large, but the superposition between the dead and alive states is decohered and the vital status of the cat is definite. This complication was known to Fr{\"o}wis and D{\"u}r\cite{Frowis2012a}, so they constructed the relative Fisher information to identify macroscopic superposition.
The relative Fisher information involves finding $N_{eff}$ for the dead and alive states, and therefore requires a complicated projective measurement that may not be experimentally feasible.

\section{Exploring indefiniteness for a cat state entangled with an auxiliary qubit}

In the above sections, we proposed, motivated and tested a new measure for macroscopicity and indefiniteness on the Gordon-Savage cat state.
The choice of the Gordon-Savage cat was made due to its potential relevance for ultra-cold atom experiment.
However, we note that the strategy of combining the extensive difference $\Lambda$ with the quality of indefiniteness $r_q$ is applicable to a broader class of cat states.
We demonstrate this by considering a cat state which is conceptually more similar to Schr\"{o}dinger's cat in which the dead and alive states are entangled with the decayed and excited states of a radioactive source.

Here, we imagine a cat state entangled with an auxiliary qubit in which the entanglement with the qubit is controlled by a parameter $\eta$ in the following way:
\begin{eqnarray}
    \label{eq:partCatBit}
    \frac{1}{\sqrt{2}}\left\{\left|a\right>\left|\uparrow\right> + \left|d\right>\left[ \left|\uparrow\right>\cos(\eta) + \left|\downarrow\right>\sin(\eta) \right]\right\}
\end{eqnarray}
where $\left|a\right>$ and $\left|d\right>$ are the dead and alive states, and are assumed to be 1) symmetric such that $\left<j_z|a\right>=\left<-j_z|d\right>$ and 2) orthogonal with respect to the identity and $J_z$: $\left<a|d\right>=\left<a|J_z|d\right>=0$.
In the limit $\cos(\eta)=1$, tracing out the qubit results in an indefinite, pure, cat-state, while in the opposite limit, $\cos(\eta)=0$, the trace results in a classical ensemble of definite alive and dead states.

Performing the analysis of indefiniteness discussed in the previous section, we compute the QFI.
While the state in Eq.~\ref{eq:partCatBit} is a pure state, we suppose we do not have access to the qubit and can only perform measurements on the cat's Hilbert space.
Therefore, we must trace out the qubit and use the general formula for the QFI of mixed states\cite{caves1994, Toth2014}:
\begin{eqnarray}
    F_{q}[\rho,\epsilon J_z]=2\sum_{l,l'}\frac{(p_{l}-p_{l'})^{2}}{p_{l}+p_{l'}}\left|\left<l\right|J_z\left|l'\right>\right|^2
\end{eqnarray}
where $\left| l\right>$ and $p_l$ are the eigenvectors and eigenvalues of the reduced density matrix respectively.
Using this expression, one obtains the QFI (see Appendix \ref{apdx:qfiqbit}) as:
\begin{eqnarray}
    F_q(\rho_{\psi=0},J_{z})=\Lambda(J_z)^2\cos^2(\eta)+PW^{2}
\end{eqnarray}
and a reduced extensive difference as:
\begin{eqnarray}
    \label{eq:catbitred}
    \Lambda r_{q}=\Lambda\sqrt{\frac{1+\alpha^2\cos^2(\eta)}{1+\alpha^2}}
\end{eqnarray}
where $PW$ is the peak width of the dead or alive(assumed to be the same) states: $PW=2\sqrt{\left<a\right|J_z^2\left|a\right>-\left<a\right|J_z\left|a\right>^2}$, and $\alpha=\frac{\Lambda}{PW}>1$ is the ratio of the extensive difference to the peak width.

Here we see that when the cat and qubit are not entangled, the quality of indefiniteness, $r_q$, quantifies a phenomenon of perfect indefiniteness, $r_q=1$, and when it is partially entangled there is imperfect indefiniteness, $r_q<1$.
In Section ~\ref{qfisec}, we argued that when $r_q<1$, and not too small, a state can be classified as indefinite if the reduced extensive difference is greater than the peak width.
For the state in Eq.~\ref{eq:partCatBit}, we find this to be the case when $\cos(\eta)>\cos(\eta_c)=\alpha^{-2}$.
If an experiment can provide a good bound using the CFI and $r_q$ is ``significant'', it will observe a quantum to classical crossover when $\eta\approx\eta_c$, in which the indefinite vital status of the cat becomes definite.

We may now consider the approximate location of the crossover, $\eta_c=\text{acos}(\alpha^{-2})$ in two limits: 1) when $\alpha=O(1)$ and 2) when $\alpha>>1$.
In the first limit,  the crossover occurs for arbitrarily small values of $\eta$ as $\alpha\rightarrow 1$.
Comparing with the Leggett-Garg experiment discussed in Appendix \ref{apdx:qfiqbit}, the Leggett-Garg inequality is violated for $\cos(\eta)>\frac{2}{3}$.
Therefore, the Leggett-Garg experiment is better at detecting the indefiniteness of the partially entangled state for $\alpha<\sqrt{\frac{3}{2}}$.
This implies that, in this limit, the projective measurement onto a dead or alive cat done in a Leggett-Garg experiment obtains more information about the mixed cat state than the Fisher Information measurement does. 

The opposite is true when $\alpha>>1$: by making $\alpha$ arbitrarily large, we can push the approximate location of the crossover to an arbitrarily amount of entanglement with the auxiliary qubit.
To make sense of this result we consider a thought experiment where the auxiliary qubit is measured and the result ignored before performing the sensitivity analysis.
In the strongly entangled limit, $\cos(\eta)<<1$, the result of this measurement is to produce a dead state $50\%$ of the time and a superposition state, $\left|\psi_s\right>\approx\left|a\right>+\cos(\eta)\left|d\right>$, in which the amplitude for the dead state is small with $\left<d|\psi_s\right>\approx\cos(\eta)$, the other $50\%$ of the time.
In this limit, a simple application of the indefiniteness condition $\Lambda r_q>PW$ suggest that this method is capable of detecting a phenomenon of indefiniteness even when the superposition produced has very little amplitude in the dead state.
A more careful consideration would note that the quality of indefiniteness is unreasonably small (not $O(0.1)$ as discussed in Section~\ref{qfisec}) and its ability to restrict the possible state which could make up a representative ensemble is severely limited.

Therefore, as noted above, we must set a bound on the quality of indefiniteness.
One way to get an intuition at what such a bound might be, is by analogy to this large $\alpha$ cat state entangled with an auxiliary qubit.
If we specify that we are only confident of a phenomenon of indefiniteness when the amplitude of the dead cat in the superposition state, $\left<d|\psi_s\right>\approx\cos(\eta)\approx r_q$, is greater than $0.1$, then we can set the threshold as $r_q>0.1$.
One could also set a more conservative threshold on the quality of indefinites by comparison with the Leggett-Garg experiment in Appendix~\ref{apdx:qfiqbit}.
There, the Leggett-Garg experiment would fail to witness indefiniteness when $\cos(\eta)=\frac{2}{3}$ and our analogous bound would be $r_q>\frac{2}{3}$.

\section{Conclusion and Discussion}
We have examined how the standard interferometric process can be used to quantify the indefiniteness of cats produced by the two-mode Hamiltonian Eq.~\ref{eq:ham}.
First, we showed that states with a large extensive difference can be produced for high temperatures initial states.
This allows an experimenter to prepare a state which, similar to Schr\"{o}dinger's cat, has uncertainty between two macroscopically different states without worrying about coherence.
We then described a possible experiment to determine the source of this uncertainty and quantify the quality of indefiniteness.
We showed how the results of this experiment can be used to infer the possible form of the pure states which could make up a possible density matrix ensemble.
This turned out to be particularly useful when describing a quantum to classical crossover where the indefinite superposition of a cat, in two macroscopically distinct states, undergoes a crossover to the definite occupation of either dead or alive. { We then finished by demonstrating the general applicability of the method to a model for which the quantum to classical crossover is controlled by the amount of entanglement with an auxiliary qubit.}

    The experiment described above involves bounding the QFI by the experimentally observable CFI {and is thus fallible to the same loopholes other Fisher Information based methods are.
    In general, these loopholes can not be tightened in the same way loopholes in Bell experiment can because there is no assumption of causally separated events:  events in an experiment that measure Fisher information could feasibly affect each other without violating special relativity.
    Instead one must make reasonable assumptions based on previous experiments, a control experiment, or a comparison with simulation.

    For example, in the bosonic interferometer experiment described above, the measurement of the CFI relies on the assumption that the Hamiltonian during the phase encoding process (step 2) is proportional to the single particle Hamiltonian encoding the phase ($J_z$ in the example considered in this paper).
    If this assumption was violated and the dynamics during the phase encoding process were highly non-linear ({\it e.g.} $J_z^4, J_x^8$), a stronger response, mimicking the effects of an indefinite state, could be observed in the distribution $p(r,\psi,\Omega)$.
    This assumption can not be checked by a causality type argument, but instead must rely on comparison with simulation or the consistency of previous experiments using bosonic interferometer.
    Without the assumption of linearity, the results of high precision measurements\cite{Casella2017, de_Angelis_2008} that use the same interferometers could not be accepted.
    One could also check the assumption of linearity by directly simulating, as done above, the predicted change in distributions $p(r,\psi,\Omega)$ and comparing with the experimental distributions.
    The tighter they match, the harder it would be to come up with a non-linear Hamiltonian that reproduces the exact same $p(r,\psi,\Omega)$.
    These simulations would also verify the assumptions made during the interferometry steps (3) and (4) after the phase has been encoded and in which further loopholes may occur.

    While simulations and references to previous experiments do not rule out peculiar possibilities in the same way the assumption of causally separated events does for Bell experiments, they do make it hard to imagine simple explanations alternative to the given assumptions.}
    Thus, the combined observation of a high quality of indefiniteness ($r_q\approx 1$) and a double peak distribution provides reasonable evidence that a cat state, which could violate a Leggett-Garg inequality, is produced by the apparatus. In addition, these measures can be acquired with current cold-atom technology and avoids the complications of the other measures discussed above.

The interpretation of the reduced extensive difference $\Lambda(J_z)r_q(J_z)$ and the arguments inferring the form of the pure states which could make up a representative density matrix ensemble can also be questioned when $r_q$ is small.
If $r_q$ is measured very close to one, then the observation of the probability distributions in Fig.~\ref{fig:probs} can be interpreted as observing the amplitudes of a pure state because $r_q$ is equal to $1$ only for pure states. 
On the other hand, when $r_q<1$, it is a qualitative judgement when comparing $\Lambda r_q$ with the peak width.
{In Section IV, we discussed one possible way to make such a qualitative judgement, but it may be interesting for future work to more rigorously investigate to what extent the combined observation of $r_q$ and the probability distribution $P(j_z)$ limit the possible states in a density matrix ensemble.
Such future work may find it useful to consider the relationship between the QFI and the resource theory of quantum invasiveness\cite{moreira2019} which is closely connected to violations of the Leggett-Garg inequalities.}
Future work will also include a study of the effects of a thermal bath and loss mechanism to identify requirements on loss, tunnelling and interaction rates for producing a cat state.

\textbf{Acknowledgements:} This work was supported in part by the NSF under Grant No. DMR-1411345, S. P. K. acknowledges financial support from the UC Office of the President through the UC Laboratory Fees Research Program, Award Number LGF-17- 476883.
The research of E. T. in the work presented in this manuscript was supported by the Laboratory Directed Research and Development program of Los Alamos National Laboratory under project number 20180045DR.

\noindent Los Alamos National Laboratory is managed by Triad National Security,
LLC, for the National Nuclear Security Administration of the U.S.
Department of Energy under Contract No. 89233218CNA000001

\appendix
\section{Cat States and Measure by  Fr{\"o}wis and D{\"u}r}
In this section we discuss the subtleties of using the measure by  Fr{\"o}wis and D{\"u}r, $N_{eff}$.
As defined in Eq.~\ref{eq:neff},  $N_{eff}$ is defined by maximizing the quantum Fisher information over all single particle generators of the phase encoding step 2 (labelled by $\Omega$ in Eq.~\ref{eq:neff}).
A naive application of this formula may lead to a wrong assessment of the indefiniteness of cat's vital status.
This is because the indefiniteness of the cat's vital status is in a specific observable ($J_z$ above), and the state of the system could have a larger QFI for a different observable.
If the conclusions where drawn directly from $N_{eff}$ one may mistakenly conclude the vital status of the cat is indefinite, while, in fact, it is a different property of the cat that is indefinite.

This possibility is manifested in the Gordon-Savage cat discussed in this paper. In Fig.~\ref{fig:QFIopt} we have plotted the QFI for all single particle observables labelled by $\Omega=(\phi,\theta)$.  Here we see that the QFI is maximum for spin pointing in the xy-plane. 
$F_q(J_z)$ still indicates that the cat is indefinite, but if one where to measure $N_{eff}$ they would observe the sensitivity to rotations around a vector perpendicular to $J_z$ ($J_y$ for the $\pi$ state), and it would tell them nothing about the indefiniteness of the cat's vital status.
It would instead tell them they had a macroscopic quantum state, but the macroscopic indefiniteness would not be in a property with clear dead and alive states distinguishable.

\begin{figure}
    \includegraphics[trim={0.5cm 0cm 0 0cm}, clip, width=0.51\textwidth]{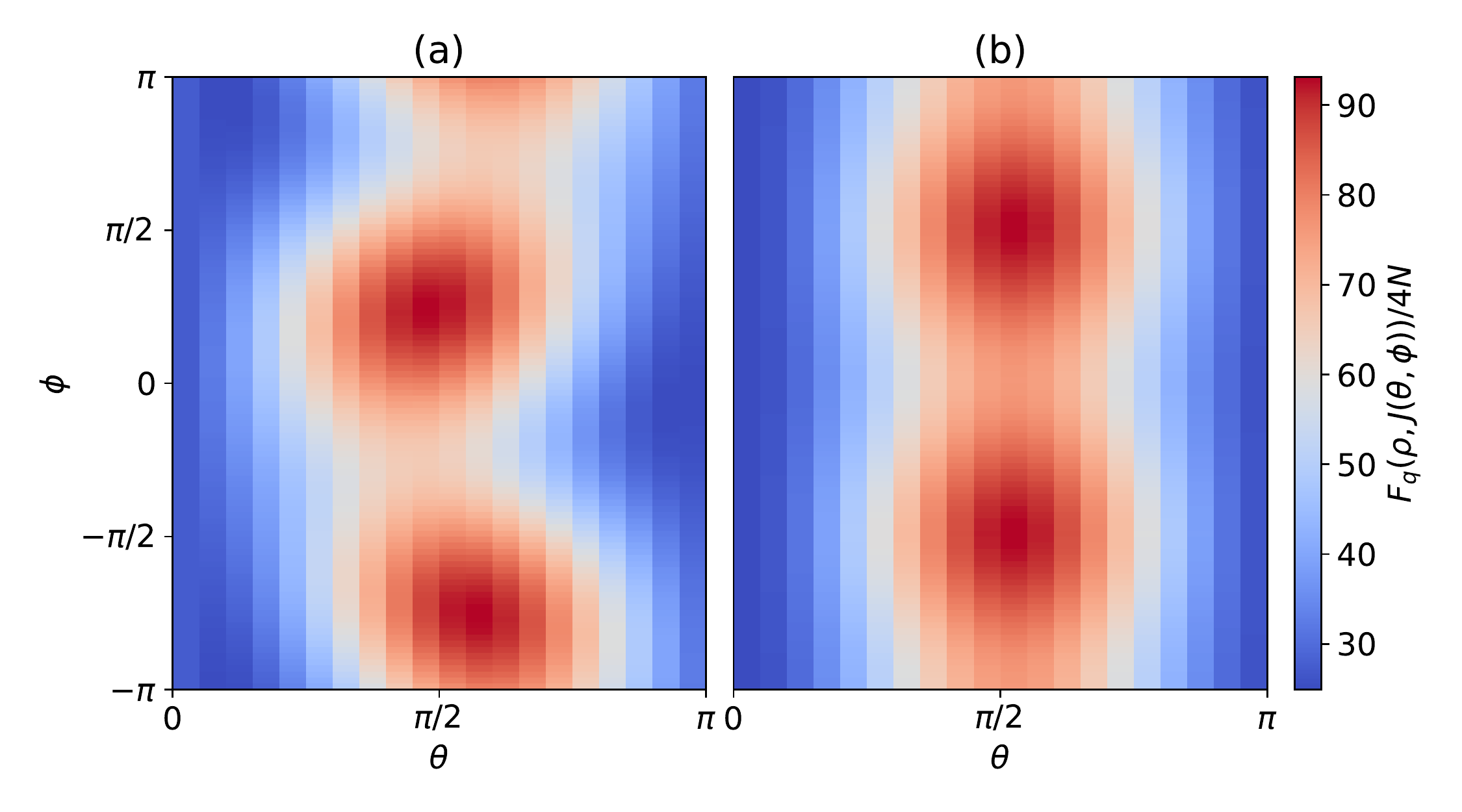}
    \caption{This is $F_{q}(\rho,J(\theta,\phi))/4N$for the $0$(Left) and $\pi$ (Right) cat states.}
    \label{fig:QFIopt}
\end{figure}

To see how this arises, we consider the Wigner distribution of the $0$ and $\pi$ cats.,
The Wigner distribution, $W(z,\phi)$, is the quasi-probability distribution function representing a quantum state, $\left|\psi\right>$:
\begin{equation}
    W(z,\phi)=\sum_{n}^{N\frac{1-z}{4}}e^{i\phi n} \left<\right. N\frac{z+1}{2}+2n\left|\psi\right>\left<\psi\right|N\frac{z+1}{2}\left.\right>,
\end{equation}
where $|N\frac{z+1}{2}\left.\right>$ are the Fock states $\left|m_1,m_2\right>$ with $m_1=N\frac{z+1}{2}$ $m_2=N-m_1$.
    The Wigner distribution has the useful property that the partial integration of one variable gives the probability distribution for the other (e.g. $P(z)=\frac{1}{2\pi}\int_{-pi}^{\pi}d\phi W(z,\phi)$) .
By considering the Wigner distributions for the $0$ and $\pi$ states (see  Fig.~\ref{fig:optWig}), we can understand the structure of the quantum state and why $N_{eff}$ may give misleading results.
The probability distributions $P(z)$ shown at the bottom of the figures indicate that the two bright red lines highlight what might be called the dead and alive cats.
The red lines individually have macroscopic uncertainty in $\phi$ and thus the xy-plane.
This implies that the dead and alive cat states are individually macroscopic quantum states.

One can now easily imagine a situation where the coherence between the dead and alive states is lost, but the dead and alive states themselves still have a large value for $N_{eff}$.
Thus if an experiment measured $F_q(J_x)$, it would find the macroscopic indefiniteness of the dead or alive cats.
One might then wrongly conclude that the vital status of the cat is indefinite when it is not.  
This complication was known to Fr{\"o}wis and D{\"u}r\cite{Frowis2012a}, so they constructed the relative Fisher information to identify macroscopic superposition.
The relative Fisher information involves finding $N_{eff}$ for the dead and alive states, and therefore requires a complicated projective measurement that may not be experimentally feasible.

\begin{figure}
    \centering
    \includegraphics[trim={0.5cm 0cm 0 0cm}, clip, width=0.485\textwidth]{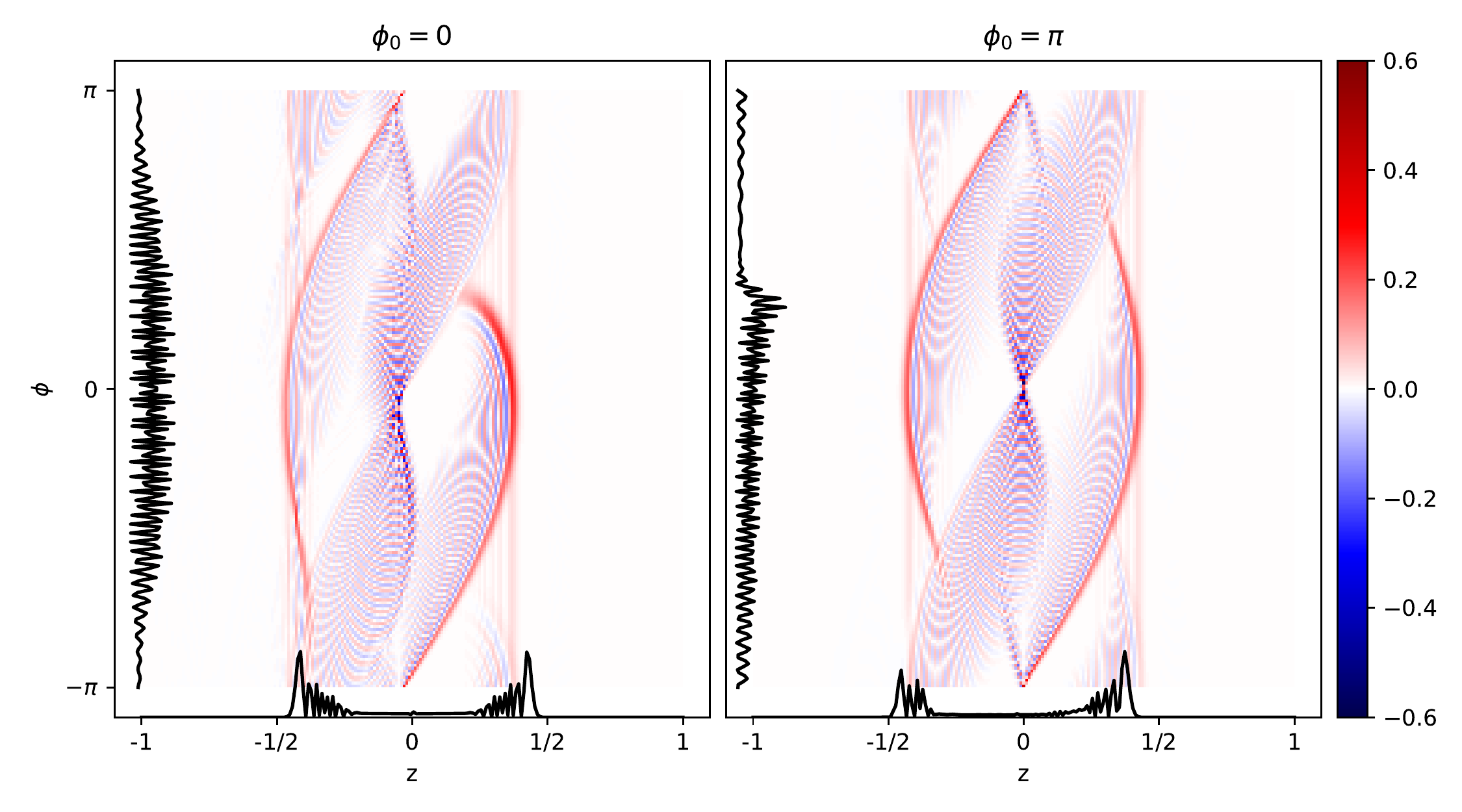}
    \caption{These are the Wigner distributions and the probabilities, $P(z)$ and $P(\phi)$, (black lines) for the $0$(Left) and $\pi$ (Right) cat states.}
    \label{fig:optWig}
\end{figure}

\label{apdx:flasePositive}

\section{Fisher Information for a Cat entangled with a qubit}
In this section we derive the expressions for the QFI of a cat entangled with a qubit discussed in the text.
The cat state entangled with a qubit is written as:
\begin{eqnarray}
    \frac{1}{\sqrt{2}}\left(\left|a\right>\left|\uparrow\right> + \left|d\right>\left( \left|\uparrow\right>\cos(\eta) + \left|\downarrow\right>\sin(\eta) \right)\right)
\end{eqnarray}
the dead and alive states are assumed to be 1) symmetric such that $\left<j_z|a\right>=\left<-j_z|d\right>$ and 2) orthogonal with respect to the identity and $J_z$: $\left<a|d\right>=\left<a|J_z|d\right>=0$.
These two assumptions imply $\left<a|J_z|a\right>=-\left<d|J_z|d\right>$ and $\left<a|J_z^2|a\right>=\left<d|J_z^2|d\right>$.
From these assumptions we derive a relationship between the variance of a (anti-)symmetric cat state, the extensive difference and the peak width as:
\begin{eqnarray}
    \label{eq:triangle}
    PW^{2}+\Lambda^2=4\left<c_{\pm}\right|J_{z}^{2}\left|c_{\pm}\right>
\end{eqnarray}
and can also write the extensive difference as:
\begin{eqnarray}
    \label{eq:transed}
    \Lambda^2=4\left|\left<c_{\pm}\right|J_{z}\left|c_{\mp}\right>\right|^2.
\end{eqnarray}

We can then derive the QFI from the following expression\cite{caves1994, Toth2014}:
\begin{eqnarray}
    F_{q}[\rho,\epsilon J_z]=2\sum_{l,l'}\frac{(p_{l}-p_{l'})^{2}}{p_{l}+p_{l'}}\left|\left<l\right|J_z\left|l'\right>\right|^2
\end{eqnarray}
where $\left| l\right>$ are the eigenvectors of the reduced density matrix and $p_{l}$ are the eigenvalues.
When tracing out the qubit we get two non-zero eigenvalues as $\frac{1}{2}(1\pm\cos(\eta))$ which we will label $l=\pm$ for the symmetric and anti-symmetric cat states and $N-1$ zero eigenvalues for the spin states orthogonal to the two cat states.
If $l=\pm$ and $l'=\mp$ the sum yields $\cos^2(\eta)\left|\left<c_{\pm}\right|J_{z}\left|c_{\mp}\right>\right|^2$.
If $l=\pm$ and $l'\neq\pm$ we can insert an identity and obtain $\frac{1\pm\cos(\eta)}{2}(\left<J_{z}^2\right>_{\pm}-\left|\left<c_{\pm}\right|J_{z}\left|c_{\mp}\right>\right|^2$.

Putting everything together with Eq.~\ref{eq:transed} and Eq.~\ref{eq:triangle}: we get 
\begin{eqnarray}
    F(J_{z})=\Lambda(J_z)^2\cos^2(\eta)+PW^{2}
\end{eqnarray}
This gives us an $r_q$:
\begin{eqnarray}
    r_{q}^2=\frac{\Lambda(J_z)^2\cos^2(\eta)+PW^{2}}{PW^2+\Lambda^2}
\end{eqnarray}

\section{Leggett-Garg violation of a cat state entangled with a qubit}
We imagine an Leggett-Garg experiment in which an initial state is evolved with respect to a Hamiltonian $\frac{H}{\hbar}=\left|a\right>\left<a\right|-\left|d\right>\left<d\right|$, and a measurement of whether the cat is alive or dead is made at $t_{1}=0$, $t_{2}=\frac{2\pi}{3}$, and $t_{3}=\frac{4\pi}{3}$.
From these measurements, correlation functions of the form $K_{ij}=\left<H_{i}H_{j}\right>$  are calculated and if the inequality:
\begin{eqnarray}
    1+K_{12}+K_{23}+K_{13}>0
\end{eqnarray}
is violated then the state must have been indefinite at some time between $t=t_1$ and $t=t_3$\cite{Leggett1985}.
If the initial state is the symmetric cat, $\frac{1}{\sqrt{2}}(\left|a\right>+\left|d\right>)$, then the violation is $-0.5$, while if the initial state is the partially entangled state in Eq.~\ref{eq:partCatBit}, the violation is $1-\frac{3}{2}\cos(\eta)$.
Thus the Leggett-Garg experiment is not capable of witnessing the indefiniteness of the entangled state for $\cos(\eta)<2/3$.

\label{apdx:qfiqbit}

\bibliography{ZQFI_DoubleWell}

\end{document}